\documentclass[12pt,epsf,amstex]{article}
\usepackage{graphicx}
\usepackage{amsmath}
\usepackage{amssymb}
\usepackage{xspace}
\usepackage{verbatim}

\addtocounter{secnumdepth}{1}
\begin{document}

\newcommand{\br}{\bar{r}}
\newcommand{\bR}{{\bf{R}}}
\newcommand{\bS}{{\bf{S}}}
\newcommand{\half}{\frac{1}{2}}
\newcommand{\bsp}{\mathbf{p}}
\newcommand{\bsA}{\mathbf{A}}
\newcommand{\bsE}{\mathbf{E}}
\newcommand{\bsL}{\mathbf{L}}
\newcommand{\bsM}{\mathbf{M}}
\newcommand{\bsV}{\mathbf{V}}
\newcommand{\bsZ}{\hat{\mathbf{Z}}}
\newcommand{\bse}{\mbox{\bf{1}}}

\newcommand{\dd}{\mbox{d}}
\newcommand{\ee}{\mbox{e}}
\newcommand{\p}{\partial}

\newcommand{\bn}{\bar{n}}
\newcommand{\bN}{\bar{N}}
\newcommand{\cL}{\cal L}
\newcommand{\cW}{\cal W}
\newcommand{\ts}{\tilde{\sigma}}
\newcommand{\tD}{\tilde{D}}

\newcommand{\la}{\langle}
\newcommand{\ra}{\rangle}

\newcommand{\OM}{OBA\xspace}
\newcommand{\XX}{SS-OBA\xspace}
\newcommand{\SPA}{single-site approximation\xspace}

\newcommand{\beq}{\begin{equation}}
\newcommand{\eeq}{\end{equation}}
\newcommand{\bea}{\begin{eqnarray}}
\newcommand{\eea}{\end{eqnarray}}
\def\lsim{\:\raisebox{-0.5ex}{$\stackrel{\textstyle<}{\sim}$}\:}
\def\gsim{\:\raisebox{-0.5ex}{$\stackrel{\textstyle>}{\sim}$}\:}

\numberwithin{equation}{section}

\thispagestyle{empty}
\title{{\Large {\bf Single-site approximation\\[2mm]
for reaction-diffusion processes\\
\phantom{xxx} }}}
 
\author{{\bf L. Canet$^1$ and H.\,J. Hilhorst$^2$}\\[5mm]
{\small $^1$ Service de Physique de l'\'Etat Condens\'e, Orme des Merisiers --
  CEA Saclay}\\[-1mm] 
{\small 91191 Gif-sur-Yvette Cedex, France}\\
{\small $^2$ Laboratoire de Physique Th\'eorique$^*$,
B\^atiment 210, Universit\'e de Paris-Sud}\\[-1mm]
{\small 91405 Orsay Cedex, France}\\}

\maketitle
\begin{small}
\begin{abstract}
\noindent 

We consider the branching and annihilating random walk 
$A\to 2A$ and $2A\to 0$ with reaction rates  $\sigma$ and $\lambda$,
respectively, and hopping rate  $D$, and study the phase diagram in the
$(\lambda/D,\sigma/D)$ plane.
According to standard mean-field theory,  
this system is in an active state for all $\sigma/D>0$,
and perturbative renormalization  
suggests that this mean-field result is valid for $d >2$;
however, nonperturbative renormalization
predicts that for all $d$ there is
a phase transition line to an absorbing state in the
$(\lambda/D,\sigma/D)$ plane. 
We show here that a simple \SPA
reproduces with minimal effort the nonperturbative phase diagram 
both qualitatively and quantitatively for all dimensions $d>2$.
We expect the approach to be useful for other reaction-diffusion
processes involving absorbing state transitions.\\

\end{abstract}
\end{small}

\noindent LPT -- ORSAY 06/04 \hspace{56mm} \\
{\small $^*$Laboratoire associ\'e au Centre National de la
Recherche Scientifique - UMR 8627}
\newpage


\section{Introduction} 
\label{secintroduction}
\vspace{5mm}

 Branching and annihilating random walks 
(BARW \cite{BramsonGray85}) have been the focus of much attention
\cite{GKT84,TT92,Jensen93,CardyTauber96,Canet04a,Canet04b},
 as they are among the simplest models of nonequilibrium
 critical phenomena  observed in physics 
 and other sciences (for  reviews see, {\it e.g.,}
 \cite{Hinrichsen00,Tauber05,Odor04}).  
 They  are generic reaction-diffusion processes in which particles
 of some species $A$ move stochastically 
on an arbitrary $d$-dimensional lattice and are subject to 
  the creation and annihilation reactions $A \xrightarrow{\sigma_m} (m+1)A$ and
  $kA \xrightarrow{\lambda_k} 0$. The nonequilibrium phase
  transitions of these models are known to belong to two distinct universality
  classes, depending on the parity of $m$ and $k$:  the ``Parity Conserving''
  class (for $m$ and $k$ even) and 
 the ``Directed Percolation'' class (for $m$ odd)
 \cite{CardyTauber96,Hinrichsen00}. 
  We are interested here in BARW belonging to 
the Directed Percolation class, 
which are generally denoted ``odd-BARW''.

The simplest odd-BARW, which in this work
we will call for short the ``\OM model'' (odd
  branching and annihilating walks), is defined by
\beq
\mbox{{\OM model}}\qquad
\left\{
\begin{array}{ll}
A \xrightarrow{\sigma} 2A \\[3mm]
2A \xrightarrow{\lambda} 0 \\[3mm]
A \emptyset \xrightarrow{D} \emptyset A
\end{array}
\right.
\label{defBAmodel}
\eeq
where $\sigma$ and $\lambda$ are {\it on-site} creation and annihilation rates,
respectively, and $D$ is a hopping rate between adjacent sites.
If the lattice has coordination number $c$, this
means that a particle will leave its lattice site at a rate $cD$.
Since it captures the essential  
critical properties of its entire class \cite{CardyTauber96},
our focus will be specifically on the \OM  model. 
Although it is well-established that the \OM model is in 
the Directed Percolation universality class \cite{CardyTauber96}, 
it has appeared much harder to determine its phase diagram,
and this is the subject of this paper. 
\vspace{3mm}

In standard mean-field (MF) theory \cite{Hinrichsen00,CardyTauber96}
the \OM model is described by the rate equation
\beq
\frac{\p \rho}{\p t}= D\Delta\rho + \sigma \rho - \lambda \rho^2,
\label{meanfieldeq}
\eeq
where $\rho(r,t) \geq 0$ is the particle density.
For all branching ratios $\sigma \geq 0$ Eq.\,(\ref{meanfieldeq}) 
has two spatially uniform
stationary solutions, the ``absorbing state'' $\rho_0=0$ and 
the ``active state'' $\rho_*=\sigma/\lambda$.
For $\sigma > 0$ the active state is stable and is reached
exponentially fast in time from any initial state with $\rho > 0$.
For $\sigma=0$ the \OM model coincides with the pure pair annihilation process
(PA) and corresponds to a critical point in parameter space at which 
the absorbing state is reached according to the power law decay
$\rho(r,t) \sim t^{-1}$. 

On a lattice in finite dimension $d$ 
the question of the stationary states of (\ref{defBAmodel}) 
is much more difficult to answer.
For $\sigma=0$, {\it i.e.} for the PA,
perturbative renormalization group analysis has shown
\cite{Lee94,CardyTauber96}  
that 
below the critical dimension $d_c=2$ the decay to the absorbing state slows
down due to depletion zones (anticorrelations) in the spatial
density distribution and follows the power law $\rho(r,t) \sim t^{-d/2}$. 

Early simulations of the \OM model \cite{TT92,Jensen93}
showed that for $d=1$ and $d=2$  
absorbing states exist even in an interval of branching ratios $\sigma>0$,
in contradistinction to MF theory. 
This indicates, therefore, that low-dimensional fluctuations qualitatively
change the ``phase diagram'' of this system in the $(\lambda,\sigma)$ plane.

In a seminal paper, Cardy and T\"auber
\cite{CardyTauber96} formulated a field theory for general BARW,
which they analyzed by perturbative renormalization group techniques.
For the \OM model they concluded 
that in dimension $d\le 2$ a minimum branching ratio
$\sigma_{\rm c}$ is needed in order for the system to be able to sustain an
active state. For small $\lambda$ the critical value $\sigma_{\rm c}$ 
behaves as
\beq
\sigma_{\rm c} \simeq D\Big( \frac{\lambda}{2D\pi\epsilon} \Big)^{2/\epsilon},
\qquad \lambda\to 0, \qquad d=2-\epsilon.
\label{exprsigmac}
\eeq
The analysis by Cardy and T\"auber is valid for small $\lambda$
 and appears to break down for $d>2$ \cite{CardyTauber96}.
However, since  $\lambda$ becomes irrelevant above two dimensions
 and since from the PA analysis one can assume that
  fluctuations are also small for $d>2$ in the OBA model
  when $\sigma$ is small, 
 Cardy and T\"auber  argue that MF theory should be restored for $d>2$,
 that is, the system should be active for all $\sigma>0$ \cite{CardyTauber96}.

This picture was modified by an analysis due to Canet {\it et al.}
  \cite{Canet04a,Canet04b},
who employed nonperturbative renormalization group (NPRG) techniques.
They demonstrated that in all finite dimensions $d>2$ there is a threshold
$\lambda_{\rm c}$ such that

${}$\phantom{i}
(i) for $\lambda<\lambda_{\rm c}$ 
the MF result of an active state for
all $\sigma>0$ remains true, but

${}$
(ii) for $\lambda>\lambda_{\rm c}$ there is an active state only when
$\sigma>\sigma_{\rm c}(\lambda)$.

\noindent
Hence according to the work of Ref.\,\cite{Canet04a,Canet04b}
standard MF theory fails dramatically for this system.

Following this failure, the \OM model has been re-investigated \cite{Odor}
 by means of  an alternative MF type approximation, 
  namely the ``cluster MF method'', also called ``generalized
  MF method'' and originally proposed for non-equilibrium systems in
   \cite{Marro99}.
 This approach consists in considering the master equation for blocks of $N$ 
  sites and truncating  the hierarchy of  probability distributions, 
  so that it can  be solved numerically. 
%
 From cluster MF calculations of the \OM model 
 \'Odor \cite{Odor} has confirmed
 for $N>2$  the existence 
of a finite threshold $\lambda_{\rm c}>0$ above which an inactive phase 
exists. 
\vspace{3mm}

The purpose of this paper is to
set up a simple analytically tractable approximation
that correctly predicts the existence of an absorbing phase
 in all finite dimensions 
 in the appropriate domain of the parameter space.
We propose a new approach, to be called {\it \SPA}, which  
allows diffusion steps to take place only to empty sites. 
The underlying idea is 
that  in dimensions $d>2$ where the intersection 
 of two directed random walks becomes unlikely,
  the destruction mechanism that drives
the system to an absorbing phase  
  is dominated by ``on-site'' annihilation 
 of a particle with its own offspring on the same site, and that
 we may neglect the annihilation of 
particles that meet due to random diffusion.
Hence the \SPA has the nature of a tree diagram approximation
in coordinate space. 
It has the virtue that it leads to what is essentially a single-site
 problem which allows for 
analytic results to be easily obtained. 
For the \OM model
these results closely reproduce the NPRG phase diagram, 
including its dimensional dependence, for all dimensions $d>2$.

We believe that in future studies this method may find application to other
reaction-diffusion processes having absorbing states and
that it may provide essential information about
their phase diagrams 
ahead of any more sophisticated work on them. 
\vspace{3mm}

In section \ref{secUBmodel} we define 
the \SPA for the \OM model,
which for short we will refer to as the ``\XX model.''
We obtain its solution in
section \ref{secsolUBmodel}. We show in section \ref{secabsorbing} that 
the model tends to an absorbing state in a specified region of the 
$(\lambda/D, \sigma/D)$ plane whose shape we discuss in section
\ref{secanalysiscalD}.
Our results are qualitatively and quantitatively
fully consistent with the NPRG results. 
In section \ref{secdiscussion} we add various comments to the discussion.
Section \ref{secconclusion} is a brief conclusion.


\section{Single-site approximation}
\label{secUBmodel}


\subsection{Definition}
\label{secdefUBmodel}

We define the \XX model as follows.
The stochastic motion of its
particles is governed by the  rules:
 
${}$\phantom{ii}
(i) Each particle is subject to the on-site 
creation reaction $A\to 2A$ at a rate $\sigma$.

${}$\phantom{i}
(ii) Each pair of particles on the same site is
subject to the annihilation 
reaction $2A\to 0$ at a rate $\lambda$.\\
\noindent
These two rules are therefore the same as in the original \OM model. 

${}$\phantom{}
(iii) Each particle may hop away from its site
at a rate $cD$ and always arrives on an empty lattice site (of some abstract
lattice that need not be specified -- say for instance to one of the next nearest 
 neighbours  if all neighbouring sites are occupied).\\ 
This rule differs from the corresponding one in the \OM model; it means that
in the \XX model the notion of lattice structure is lost.  

For $D=0$ the two models
are identical; 
we therefore expect the \XX model to be a good approximation of
the \OM model in the small diffusion regime.


\subsection{Solution}
\label{secsolUBmodel}

To see that the \XX model is exactly soluble,
it suffices to note that no particle ever enters an occupied site from the
outside 
and that therefore each active site
has a dynamics independent of the others;
a site's occupation number evolves only due to on-site
creation and annihilation transitions and to
departures.
The solution therefore decomposes into the analysis of the time evolution on a
single 
site given its initial condition at some time $t_0$,
and the analysis of the coupling between sites due to the diffusion mechanism.
These two questions are studied in subsections \ref{secsinglesite} and
\ref{seccoupling}, respectively.


\subsubsection{Single-site problem}
\label{secsinglesite}

Let $P(n,t)$ be the probability that a specific site
contains exactly $n$ particles at time $t$.
Then this probability satisfies the master equation
\begin{eqnarray}
\frac{\rm{d}}{{\rm d} t}P(n,t) &=& 
 \sigma(n-1)P(n-1,t)-\sigma n P(n,t)\nonumber\\[0.5mm]
&& +\mbox{$\frac{1}{2}$}\lambda(n+1)(n+2)P(n+2,t)
-\mbox{$\frac{1}{2}$}\lambda n(n-1)P(n,t)\nonumber\\[2mm]
&& + cD (n+1) P(n+1,t) - cD n P(n,t)
\label{mastereq}
\end{eqnarray}
for $n=0,1,2,\ldots$ and with the convention that $P(-1,t) \equiv 0$.
Here the lattice coordination number $c$ is the only parameter
reminiscent of the original lattice.
We introduce the scaled variables 
\beq
\tau=\lambda t, \qquad \tilde{\sigma}=\sigma\lambda^{-1}, \qquad
\tilde{D} = cD\lambda^{-1}
\label{deftildesD}
\eeq
and set $P(n,t)={p}(n,\tau)$. 
Defining the vector $\bsp(\tau)=({p}(0,\tau),{p}(1,\tau),\ldots)$ 
we may write (\ref{mastereq})  as
\beq
\frac{\dd\bsp}{\dd\tau} = {\bsM}{\bsp},
\label{eqnPM}
\eeq
where $\bsM$ is the tetradiagonal matrix
\beq
\bsM = 
\left(
\begin{array}{cccccc}
0 &      \tD &            1  &            0 &            0 & \ldots \\[2mm]
0 & -\tD-\ts &         2\tD  &            3 &            0 & \ldots \\[2mm]
0 &      \ts & -2\tD-2\ts-1  &         3\tD &            6 & \ldots \\[2mm]
0 &        0 &         2\ts  & -3\tD-3\ts-3 &         4\tD & \ldots \\[2mm]
0 &        0 &            0  &         3\ts & -4\tD-4\ts-6 & \ldots \\[2mm]
\ldots&\ldots&       \ldots  &       \ldots &       \ldots & \ldots
\end{array}
\right).\\[2mm]
\label{defM}
\eeq
Eqs.\,(\ref{eqnPM})--(\ref{defM}) 
constitute a problem with two parameters, $\tD$ and $\ts$.
Except when $\tilde{\sigma}=\tilde{D}=0$ we expect Eq.\,(\ref{eqnPM})
to have only a single stationary solution, {\it viz.}
${p}^{\rm st}(n)=\delta_{n,0}$. The reason is that as $n$ gets large, the
annihilation rate dominates the creation by one order in $n$, which prevents 
``escape'' of the site occupation number to $n=\infty$; 
the occupation number, therefore, can get caught only in $n=0$, 
even though for large $\tilde{\sigma}$
it may be in a long-lived ``metastable'' state.
When $\tilde{\sigma}=\tilde{D}=0$ the particle number can change only
by pair annihilation, which is a parity conserving process. 
There will then be two independent stationary states, {\it viz.}
${p}_0^{\rm st}(n)=\delta_{n,0}$ and
${p}_1^{\rm st}(n)=\delta_{n,1}$.

We will write $G_{nn_0}(\tau-\tau_0)$, where $\tau-\tau_0 \geq 0$ and
$n, n_0=0,1,2,\ldots$, 
for the solution of Eq.\,(\ref{eqnPM})
with initial condition ${p}(n,\tau_0)=\delta_{nn_0}$.
It is not possible in the general case
to write this solution in an explicit closed form, but
we will suppose that all its essential properties 
can be determined.
In particular we assume that, 
except when $\tilde{\sigma}=\tilde{D}=0$, 
the function
$G_{nn'}(\tau)$ tends to zero exponentially at some time scale 
$1/\mu_1(\tilde{\sigma},\tilde{D})$,
\begin{equation}
G_{nn'}(\tau)\sim {\rm e}^{-\mu_1\tau}, \qquad \tau\to\infty,
\qquad (\ts,\tD) \neq (0,0), \qquad n,n'>0,
\label{defkappa}
\end{equation}
with $\mu_1( \tilde{\sigma},\tilde{D} ) > 0$.
When $\tilde{\sigma}=\tilde{D}=0$ we have $\mu_1(0,0)=0$,
which signals the degeneracy of the stationary state. 
This completes the discussion of the single site problem.


\subsubsection{Coupling between sites}
\label{seccoupling}

Let at time $\tau=0$
the initial state be such that there are $S_n(0)$
sites with occupation number $n$, where $n=1,2\dots$. 
The total initial particle number $N(0)$ is then given by
\begin{equation}
N(0) = \sum_{n=1}^\infty n S_n(0). 
\label{relNS}
\end{equation}
We are interested in the average number $\langle N(\tau)\rangle$ 
of particles at some arbitrary instant of time $\tau>0$;
here $\la\ldots\ra$ denotes an average with respect to 
the initial distribution of the $S_n(0)$ 
and the stochastic time evolution. We will proceed by first calculating the 
averages $\langle S_n(\tau)\rangle$.

There are two types of sites, those that 
are occupied initially, and those that 
get occupied only later during the time evolution due to diffusion steps.
We denote the contribution of these two types of sites
by superscripts  $(0)$ and  $(1)$, respectively, so that
\begin{equation}
\langle S_n(\tau)\rangle = \langle S_n(\tau)\rangle^{(0)}
\,+\,\langle S_n(\tau)\rangle^{(1)}.  
\label{exprSSS}
\end{equation}  
Upon considering the time evolution of the initially occupied sites we
find
\begin{equation}
\langle S_n(\tau)\rangle^{(0)} = \sum_{n'=1}^\infty
G_{nn'}(\tau)S_{n'}(0). 
\label{exprS1}
\end{equation}
Throughout the time interval $0<\tau'<\tau$
new occupied sites are created due to diffusion steps
at a rate $\tilde{D}\langle N(\tau')\rangle$.
Hence
\begin{equation}
\langle S_n(\tau)\rangle^{(1)} = \tilde{D}\,\int_0^\infty {\rm d}\tau'\,
G_{n1}(\tau-\tau') \langle N(\tau')\rangle,
\label{exprS2}
\end{equation}
where the upper boundary of the integral has been sent to $\infty$ 
 exploiting  that $G_{n n'}(t)$ vanishes for $t<0$.
Summing Eqs.\,(\ref{exprS1}) and (\ref{exprS2}) 
yields
\begin{eqnarray}
\langle S_n(\tau)\rangle &=&
\sum_{n'=1}^\infty G_{nn'}(\tau)S_{n'}(0)\nonumber\\
&& +\, \tilde{D}\,\int_0^\infty\,\, {\rm d}\tau'\,
G_{n1}(\tau-\tau') \langle N(\tau')\rangle. 
\label{exprS}
\end{eqnarray}
When multiplying this equation by $n$ and summing on $n$ we obtain
\begin{eqnarray}
\langle N(\tau)\rangle &=&
\sum_{n'=1}^\infty\sum_{n=1}^\infty n G_{nn'}(\tau) S_{n'}(0)\nonumber\\
&& + \, \tilde{D}\,\int_0^\infty {\rm d}\tau'
\sum_{n=1}^\infty n G_{n1}(\tau-\tau') \langle N(\tau')\rangle,
\label{eqnNav}
\end{eqnarray}
which is a closed equation for $\langle N(\tau)\rangle$. 

For convenience let us now restrict the initial states to those that
have only singly occupied sites; that is, we take 
$S_n(0) = N(0)\delta_{n1}$. 
This clearly does not affect the long time behaviour of the system so it does
not restrict  the generality of our argument.
Furthermore we define
\begin{equation}
H(\tau-\tau')=\sum_{n=1}^\infty n G_{n1}(\tau-\tau').
\label{defHt}
\end{equation}
When substituting the previous definitions in Eq.\,(\ref{eqnNav}) we find
\begin{equation}
\langle N(\tau)\rangle = N(0)H(\tau)
+ \tilde{D}\,\int_0^\infty {\rm{d}}\tau' H(\tau-\tau')
\langle N(\tau') \rangle.
\label{eqnNav1}
\end{equation}
In terms of the Laplace transforms
\begin{equation}
\hat{N}(s)=\int_0^\infty {\rm d}\tau\,\, {\rm e}^{-s\tau} \langle
N(\tau)\rangle, 
\qquad \hat{H}(s)=\int_0^\infty {\rm d}\tau\,\, {\rm e}^{-s\tau} H(\tau),
\label{defLtransf}
\end{equation} 
it becomes $\hat{N}(s)=N(0)\hat{H}(s) + 
\tilde{D}\hat{N}(s)\hat{H}(s)$,
whence the solution
\begin{equation}
\hat{N}(s)=N(0)\,\frac{\hat{H}(s)}{1-\tilde{D}\hat{H}(s)},
\label{solNav1L}
\end{equation}
which may be inverse Laplace transformed to $\langle N(\tau)\rangle$.
This completes the solution of the average
total particle number in the \XX  model.
\vspace{2mm}


\subsection{Existence of an absorbing phase}
\label{secabsorbing}

Because of Eqs.\,(\ref{defHt}) and 
(\ref{defkappa}) the decay of $H(\tau)$ will be characterized by
the same $\mu_1$ as that of $G_{nn'}(\tau)$.
Let us suppose that $H(\tau)={\rm e}^{-\mu_1\tau}$,
so that $\hat{H}(s)=1/(\mu_1+s)$.
Then Eq.\,(\ref{solNav1L}) implies that
$\hat{N}(s)=N(0)/(\mu_1-\tilde{D}+s)$, whence
\begin{equation}
\langle N(\tau)\rangle = N(0){\rm e}^{-(\mu_1-\tilde{D})\tau}.
\label{exsolN}
\end{equation}
Therefore  the condition for $\langle N(\tau)\rangle$
to tend to zero is
\begin{equation}
\mu_1\big(\tilde{\sigma},\tilde{D}\big)\,-
\,\tilde{D}\,>\,0.
\label{cond}
\end{equation}
The \XX model has an absorbing phase in the region of parameter space where
Eq.\,(\ref{cond}) holds.
\vspace{2mm}

\begin{figure}[ht]
\begin{center}
{\includegraphics[height=94mm,angle=-90]{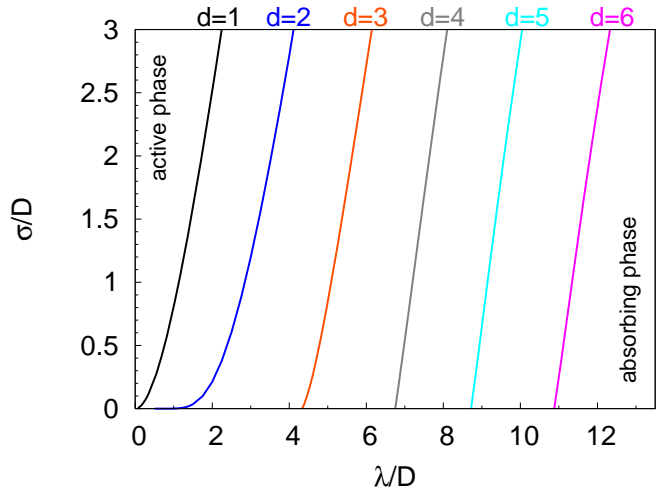}}
\caption{\small Phase diagram of the \OM model (\ref{defBAmodel})
ensuing from the nonperturbative renormalization group (NPRG) 
in dimensions $d=1,\ldots,6$, from Canet {\it et al.} \cite{Canet04b}. 
For each dimension, the active phase lies on the left and the absorbing phase
on the right of the transition line. For $d>2$ this line
comes into the $\lambda/D$ axis 
at some threshold value $\lambda_{\rm c}/D$ 
with a finite slope $\alpha_{\rm NPRG}\approx 2.3$.
The spacing between the phase transition lines for two successive dimensions
is $\Delta_{\rm NPRG}\approx 2.2$.} 
\label{figphasediagram}
\end{center}
\end{figure}

To analyze this equation we consider first the special case $\tilde{D}=0$.
In this case the \XX model is described by a single-site master equation 
identical to Eq.\,(\ref{mastereq}) but with $\tilde{D}=0$.
It reaches the absorbing state exponentially fast at an asymptotic rate
$\mu_1(\tilde{\sigma},0)$ which for $0<\tilde{\sigma}<\infty$
satisfies  
\beq
\mu_1(\tilde{\sigma},0)>0.
\label{condDzero}
\eeq
We note that $\mu_1(0,0)=0$ as emphasized in section \ref{secsinglesite}
and that we must furthermore have
$\lim_{\ts\to\infty}\mu_1(\ts,0)=0$, since
the decay of the metastable state becomes infinitely slow in that limit.
We invoke now continuity of $\mu_1(\ts,\tD)$ in its second argument 
and conclude that
for all values of $\tilde{\sigma}$ in $(0,\infty)$ there exists a positive 
threshold value ${\cal D}_{\rm c}(\tilde{\sigma})$
such that Eq.\,(\ref{cond}) is satisfied for all
$0 \leq \tilde{D} < {\cal D}_{\rm c}(\tilde{\sigma})$,
{\it i.e.} the stationary state of the \XX model is
absorbing for all 
$0 \leq \tilde{D} < {\cal D}_{\rm c}(\tilde{\sigma})$.
Reverting to the original parameters this means that for all
ratios $0<\sigma/\lambda<\infty$ there exists a 
${\cal D}_{\rm c}(\sigma/\lambda)$ such that for 
\beq
\frac{\lambda}{D}  > \frac{c} { {\cal D}_{\rm c}(\sigma/\lambda) }
\label{condorig}
\eeq
the stationary state is absorbing.  Written this way, inequality
\,(\ref{condorig})  allows for comparison with the phase 
diagrams of Ref.\,\cite{Canet04b} 
which are plotted in the plane of abscissa $\lambda/D$
and ordinate $\sigma/D$ and are displayed  in
Fig.\,\ref{figphasediagram}.
%
%
%
%
First of all 
Eq.\,(\ref{condorig}) implies  the existence for large enough $\lambda$
of  an absorbing region in the phase diagram, which 
is in full agreement with the NPRG results of Fig.\,\ref{figphasediagram}. 
Furthermore, as the spatial dimension $d$ tends to infinity, so does  the
lattice 
coordination number $c$ (typically linearly with $d$), and 
 Eq.\,(\ref{condorig}) shows that
  the region of phase space to which our proof
applies, recedes to infinity as $c\to\infty$. This, too, corroborates the
 NPRG results of Ref.\,\cite{Canet04b} which indicate that an absorbing phase
  exists in all finite dimensions.  It  also  matches with standard
 MF theory, which has effectively $d=\infty$, 
 and  no absorbing phase in this limit. 

One can take a further step and analyze the shape of the phase transition line
between 
the active and absorbing phases obtained in the \XX model. As shown in the
next section, it fits  with the NPRG results even on a quantitative level.  


\subsection{Analysis of the phase diagram}
\label{secanalysiscalD}

In this section we analyze in greater details
the location of the phase transition line
in the $(\lambda/D,\sigma/D)$ plane defined by Eq.\,(\ref{condorig}).
We show that it intersects the $\lambda/D$ axis 
at some finite value $\lambda/D=\lambda_{\rm c}/D$, and with a positive slope.
To do so we determine 
$\mu_1(\ts,\tD)$
for arbitrary $\tD$ perturbatively in small $\ts$.
To linear order in $\ts$, the result, derived in the Appendix, is
\beq
\mu_1(\ts,\tD) = \tD - \frac{\tD-1}{\tD+1}\,\ts\, + {\cal O}(\ts^2).
\label{reskappa}
\eeq 
Combining this expression with criterion (\ref{cond}) 
we see that for $\ts\to 0$,
the system will tend towards an absorbing state at the condition that
\beq
\tD<1 \qquad \mbox{or} \qquad \lambda>\lambda_{\rm c}\,,
\label{condtD}
\eeq
where
\beq
\lambda_{\rm c}/D = c = 2d,
\label{thres}
\eeq
and in which the last equality is for the case of a 
hypercubic $d$-dimensional lattice.

The variation
of the threshold $ \lambda_{\rm c}/D$ with the dimension has also been
obtained \cite{Canet04b} within NPRG.
Indeed, from Fig.\,\ref{figphasediagram}, which is for hypercubic lattices,
this variation appears to be linear,
$ \lambda_{\rm c}/D \simeq \Delta_{\rm NPRG} d $ with the slope equal to
$ \Delta_{\rm NPRG} \approx 2.2 $ (as estimated in \cite{Canet04b}).
This result is in very close agreement with our Eq.\,(\ref{thres}),
which yields $\Delta=2$ for the same slope.
%
%
\begin{figure}[ht]
\begin{center}
{\includegraphics[height=63mm]{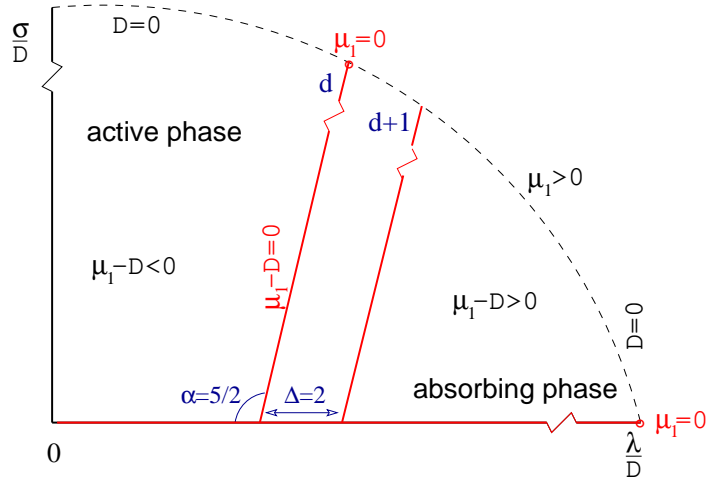}}
\caption{\small Phase diagram of the \OM model (\ref{defBAmodel}) 
according to the \SPA
 of this work, for the dimensions $d$ and $d+1$.
The circle at infinity has been represented by a dashed arc.
The phase transition lines come into the $\lambda/D$ axis with a slope
$\alpha=\frac{5}{2}$ and for two successive dimensions have a spacing
$\Delta=2$. This phase diagram is very close qualitatively and quantitatively
to the one of Fig.\,\ref{figphasediagram}.}
\label{diag}
\end{center}
\end{figure}
%
%

The second order correction in $\ts$ to  $\mu_1$ can also be worked out, 
as shown in the Appendix,  and allows for the determination of the slope  
of the transition line.
To second order the smallest eigenvalue is given by
\beq
\mu_1(\ts,\tD) = \tD - \frac{\tD-1}{\tD+1}\,\ts\, +\frac{2(2\tD^2-
  3\tD-3)}{(2\tD +3)(\tD+1)^3}\ts^2+ {\cal O}(\ts^3). 
\label{muorder2}
\eeq
According to the criterion of Eq.\,(\ref{cond})
the phase transition line between the active and the absorbing 
phases is defined by the condition $\mu_1-\tD=0$. 
Substituting (\ref{muorder2}) in this condition, dividing out $\ts$, 
writing $\ts=(\sigma/D)(\tD/c)$, and using
that $\ts$, $\tD-1$, and $\lambda/D-c$  are of the same order 
we find
\bea
\frac{\sigma}{D} &=& \frac {(2\tD+3)(\tD+1)^2} {2\tD(2\tD^2-3\tD-3)}\,
c(\tD-1)
\,+\, {\cal O} \big( (\tD-1)^2 \big) \nonumber\\[2mm]
&=& \frac{5}{2}\left( \frac{\lambda}{D}-c \right)
\,+\, {\cal O} \left( \Big(\frac{\lambda}{D}-c\Big)^2 \right),
\label{eqnseparatrix}
\eea   
which is the equation for the phase transition line near the threshold 
point $(\lambda/D,\sigma/D)=(c,0)$.

The slope $\alpha$ at the threshold appears to be $\alpha=\frac{5}{2}$,  
independently of $c$, {\it i.e.,} of the dimension $d$. 
The value $\alpha=\frac{5}{2}$ compares favorably with the NPRG phase
diagram of Fig.\,\ref{figphasediagram}, where 
the phase transition lines appear to be merely drifting as the dimension
grows, with a quasi-constant slope $\alpha_{\rm NPRG}$ numerically equal
to $\alpha_{\rm NPRG} \approx 2.3$ \cite{footnote1}.\\

Fig.\,\ref{diag} summarizes the results of this section and
shows that our \SPA captures all the essential features of
 phase diagrams of the \OM model in $d>2$. It allows in particular to probe
 the dependence on the dimension, which is beyond the scope of standard MF
 approach.


\section{Discussion}
\label{secdiscussion}

In addition to the discussion that has accompanied the above determination of
the phase diagram, several points deserve some comments. We provide them here.
\vspace{2mm}

{\it Order parameter.\,\,} Our first remark concerns the active state.
According to Eq.\,(\ref{cond}) this state is characterized by
the inequality $\mu_1-\tD<0$.  In the region of the phase diagram where
 this inequality  holds, Eq.\,(\ref{exsolN}) 
shows that the total particle number
increases exponentially in time.  Nevertheless, in this regime 
the average number of particles per occupied site,
$\la n\ra_{\rm occ}$, is well defined in the limit $\tau\to\infty$. 
One may consider this quantity as the
order parameter of the \XX, but it should be emphasized that the usual
order parameter is instead
$\rho = q\la n\ra_{\rm occ}$ where $q$ is the fraction of occupied sites.

In the \XX model each diffusion step takes the diffusing particle to an
empty site.  For the discussion of the phase diagram in the preceding sections 
there has been no need to
specify whether this is a site that has
perhaps previously been occupied or whether it is an entirely new site.
Therefore the fraction $1-q$ of empty sites, and hence the usual order
parameter $\rho$, remain undefined in the \XX model.


Hence a calculation of $\rho$ within our approach
would require further elaboration and/or modification of the model.
A way to do this was pointed out by Dickman \cite{Dickman06}, but goes
beyond our more restricted purpose of analytically studying the phase diagram.
\vspace{2mm}

{\it Reformulation.\,\,}
The \XX model may be formulated in an equivalent way 
which preserves the original lattice structure. 
This reformulation requires that we distinguish between two notions
that coincided in the definition of section (\ref{secdefUBmodel}), 
{\it viz.} between a ``lattice site'' in its usual sense and a ``family'' of
particles: 
a {\it family\,} consists of a particle having diffused to a 
(now not necessarily empty) lattice site, 
together with all the offspring it has generated on that site
and which has never left it.
Initially all particles on the same site are considered to constitute a family.
Consequently, the particles on each site may at any time
be partitioned into families. When a particle performs a diffusion step, it
leaves its family and starts a family of its own on its arrival site,
where other families may or may not already be present.

We may then replace
rules (ii) and (iii) of section \ref{secdefUBmodel} by the following:

${}$\phantom{i}
(ii$^\prime$) Each particle can annihilate (at a rate $\lambda$ per pair)
only with other members of its own family on that site. 

${}$\phantom{}
(iii$^\prime$) Each particle performs diffusion steps to neighboring sites
with a rate $D$ per transition; for coordination number $c$
this means that a particle will leave its site at a rate $cD$.

This reformulation of the \XX does not change the mathematics and, in
particular, leads to the same 
$\la N(\tau)\ra$ as found in section \ref{secsolUBmodel}. 
It has the merit of bringing out  clearly that 
two-particle annihilation in the \XX model occurs under
more restrictive conditions than in the original \OM model.
This makes it tempting to believe that the average
total particle number in the \XX model is an upper bound to the same quantity 
in the \OM model. If true, our calculation would constitute an exact proof of
the existence of an absorbing state in the OBA model. 
We have not, however, been able to prove this upper bound property and leave it
as an open problem.
\vspace{2mm}

{\it Below two dimensions.\,\,}
Our final remark concerns what happens 
when the spatial dimension $d$ is equal
to or less than the critical dimension $d_c=2$ for pure pair annihilation.
For $d \leq d_{\rm c}$ the renormalisation group (both perturbative and
 nonperturbative) predicts that in the \OM model the threshold
$\lambda_{\rm c}$ vanishes.
In the \XX model this
would correspond for instance to ${\cal D}_{\rm c}(0) \to \infty$ in
 Eq.\,(\ref{condorig}) for $d\le 2$. We have not investigated this point
 further, since the behaviour of the system for small $\lambda/D$ and
 $\sigma/D$ corresponds to the large diffusion regime, for which we do not
{\it a priori} expect the \XX model 
to be a good description of the \OM model.  


\section{Conclusion}
\label{secconclusion}

We have formulated a new approximation for
reaction-diffusion problems with an absorbing-state transition. 
Its characteristic feature is that it forbids annihilation reactions 
when one or more of the participating particles have moved from the
 site where they were originally created.
The approach then leads to what is essentially
a single-site calculation,
 which allows the consequences to be determined analytically. 
 
We have applied the approximation to the \OM model
 $A\to 2A$ and $2A\to 0$. 
We have shown that our theory produces qualitatively and quantitatively
the main properties of its phase diagram in agreement with the predictions of
nonperturbative renormalization but with far less effort. 
\vspace{4mm}


\section*{Acknowledgments}

LC wishes to thank her collaborators of Ref. \cite{Canet04a,Canet04b} on the
  NPRG calculation which inspired this work.  
  Part of this  work has benefited  from the financial support granted to LC
  by the European Community's Human Potential Programme under contract
  HPRN-CT-2002-00307, DYGLAGEMEM. 
HJH has benefited from a six month sabbatical period (CRCT) 
awarded to him by the French Ministry of Education in 2004-2005.


\appendix

\section{Appendix}

We wish to calculate  
the smallest nonzero eigenvalue, called $-\mu_1(\ts,\tD)$, 
of the matrix $\bsM$ defined by
Eq.\,(\ref{defM}). We will perform this calculation for arbitrary $\tD$ and
perturbatively for small $\ts$.
The symbol $\bse$ will denote the identity matrix.
For any matrix $\bsL$ with rows and columns labeled by $n=0,1,2,\ldots,$
we write $\bsL^{(j)}$ for the matrix obtained from it
by erasing its rows and columns of indices $n=0,1,\ldots,j-1$; 
and we denote by $[\bsL]_{mm'}$ the matrix obtained from it
by erasing the row $m$ and column $m'$.
Hence $\bsM^{(0)}=\bsM$ and $\bsM^{(1)}=[\bsM]_{00}$.

If we modify the matrix $\bsM$ by suppressing its
subdiagonal (of elements $0,\ts,2\ts,3\ts,\ldots$),
it becomes upper triangular. This modified matrix has
eigenvalues $-\nu_k$ given by its diagonal elements, {\it i.e.,}
\beq
\nu_k = k\tD +k\ts + \tfrac{1}{2}k(k-1) \equiv \nu_k^{(0)}+k\ts, \qquad
k=0,1,2,\ldots.  
\label{defmul}
\eeq
Hence its smallest nonzero eigenvalue is $\nu_1=\tD+\ts$. 
Restoring now the subdiagonal will change the $\nu_k$ 
and we will calculate this change perturbatively to second order in $\ts$. 
That is, we will look for a solution $\mu=\mu_1$
of Eq.\,(\ref{eqnmubis}) which has the form
\beq
\mu_1 = \nu_1 + \Delta\nu_1  
\label{defDeltamu1}
\eeq
with
\beq
\Delta\nu_1\equiv  \nu_1^{(1)}
\ts + \nu_1^{(2)} \ts^2 + {\cal O}(\ts^3).
\label{defDeltanu1}
\eeq
The nonzero eigenvalues $\mu$ of $\bsM$ satisfy
\beq
\det\big( \bsM^{(1)}+\mu\bse^{(1)} \big) = 0.
\label{eqnmu}
\eeq
A cofactor expansion of (\ref{eqnmu}) along the column $n=0$ gives 
\beq
-(\nu_1-\mu) \det\big( \bsM^{(2)} + \mu\bse^{(2)} \big) 
- \ts \det\big( [\bsM^{(1)} + \mu\bse^{(1)}]_{10} \big)  = 0, 
\label{eqnmubis}
\eeq
whence, after we substitute (\ref{defDeltamu1}) in (\ref{eqnmubis}),
\beq
\Delta\nu_1  = \ts\, \frac
{ \det\big( [\bsM^{(1)} + \mu_1 \bse^{(1)}]_{10} \big) }
{ \det\big( \bsM^{(2)} + \mu_1\bse^{(2)} \big) }.  
\label{eqn2}
\eeq
The first order correction $\nu_1^{(1)}$ follows from linearizing
Eq.\,(\ref{eqn2}) in $\ts$, which amounts to setting 
 $\mu_1=\nu_1^{(0)}=\tD$ and $\ts = 0$ in
  the two determinants of the right hand side of (\ref{eqn2}). 
Both determinants then reduce to a product of diagonal
elements and identical factors cancel. Using the explicit expression 
of $\bsM$ together with (\ref{defDeltanu1}) we then get from (\ref{eqn2})
the first order coefficient 
\beq
 \nu_1^{(1)} = - \frac{2 \tD}{\tD+1},
\label{rescoeff1}
\eeq
which leads to the result shown in Eq.\,(\ref{reskappa}).

The second order in $\ts$
in the expansion (\ref{defDeltanu1}) determines the slope at which
the phase transition line comes into the $\lambda/D$ axis 
in the phase diagram. 
This second order correction requires that one computes the determinant
ratio in  Eq.\,(\ref{eqn2}) 
 to linear order in $\ts$. Performing a cofactor expansion of both
 determinants  
 along their column $n=0$, one gets
\begin{eqnarray}
\det\big( [\bsM^{(1)}+ \mu_1\bse^{(1)}]_{10}\big) &=& 2\tD \det\big(\bsM^{(3)}
 + \mu_1\bse^{(3)}\big) \nonumber \\ 
&& -2\ts  \det \big([ [\bsM^{(1)}+\mu_1 \bse^{(1)}]_{10}]_{10}\big),\nonumber
 \\[2mm] 
 \det\big( \bsM^{(2)} + \mu_1\bse^{(2)} \big) 
&=& -(\nu_2-\mu_1)\det\big(\bsM^{(3)}+ \mu_1\bse^{(3)}\big)\nonumber\\ 
&& -2\ts \det \big( [\bsM^{(2)}+\mu_1\bse^{(2)}]_{10}\big).
\label{dvptdet} 
\end{eqnarray}
To obtain the expansions of Eqs.\,(\ref{dvptdet}) to first order in $\ts$, it
suffices that we evaluate the two determinantal coefficients
of $\ts$ in the second and fourth line above to zeroth order.    
Furthermore,  it turns out that the first order contributions of
$\det\big(\bsM^{(3)}+ \mu_1\bse^{(3)}\big)$ in Eqs.\,(\ref{dvptdet})  cancel
out in  
 the ratio  (\ref{eqn2}), so that only the  zeroth order of this determinant
 is actually needed as well. The zeroth order of all the determinants is again 
 obtained by  setting  $\mu_1=\nu_1^{(0)}$ and $\ts=0$, upon which the
 matrices  become   upper triangular and the determinants reduce to  the
 products of  diagonal elements.  Finally,  
\beq
\nu_1^{(2)} = \frac{2(2\tD^2- 3\tD-3)}{(2\tD +3)(\tD+1)^3}.
\label{rescoeff2}
\eeq
By combining (\ref{defDeltamu1}), (\ref{defDeltanu1}), (\ref{rescoeff1}), 
and (\ref{rescoeff2}) one obtains the second order result (\ref{muorder2})
exploited in the main text.



\begin{thebibliography}{10}
\bibitem{footnote1}
The reaction rates defined in Eq.\,(1)
of \cite{Canet04b} are such
that one has the correspondence $\sigma_{\rm NPRG}= 2 \sigma$.
Fig. \ref{figphasediagram} has been  plotted with the definition of this
work.
\end{thebibliography}
\end{document}